\documentclass{article}

\usepackage[margin=1in]{geometry}
\usepackage{amsmath,amssymb,amsthm}
\usepackage{calc}
\usepackage{graphicx}
\usepackage{longtable,booktabs,array,multirow}
\usepackage[authoryear,round]{natbib}
\usepackage{hyperref}
\usepackage{ifxetex,ifluatex}
\usepackage{setspace}

\ifnum 0\ifxetex 1\fi\ifluatex 1\fi=0
  \usepackage[T1]{fontenc}
  \usepackage[utf8]{inputenc}
  \usepackage{textcomp}
\else
  \usepackage{unicode-math}
  \defaultfontfeatures{Scale=MatchLowercase}
  \defaultfontfeatures[\rmfamily]{Ligatures=TeX,Scale=1}
\fi

\hypersetup{hidelinks,pdftitle={Equilibrium Selection and Closure in
DSGE Solution Methods}}
\newtheorem{theorem}{Theorem}[section]
\newtheorem{proposition}{Proposition}[section]
\theoremstyle{definition}

\theoremstyle{remark}

\newcommand{\nameddefinition}[2]{\noindent\refstepcounter{definition}\label{#1}\textbf{Definition \thedefinition\ (#2).}}
\newcommand{\namedaxiom}[2]{\noindent\refstepcounter{axiom}\label{#1}\textbf{Axiom \theaxiom\ (#2).}}
\newcommand{\proofheading}[1]{\par\noindent\textit{#1.}}
\newcommand{\proofend}{\hfill\ensuremath{\square}}
\providecommand{\tightlist}{%
  \setlength{\itemsep}{0pt}\setlength{\parskip}{0pt}}

\makeatletter
\def\maxwidth{\ifdim\Gin@nat@width>\linewidth\linewidth\else\Gin@nat@width\fi}
\def\maxheight{\ifdim\Gin@nat@height>\textheight\textheight\else\Gin@nat@height\fi}
\makeatother
\setkeys{Gin}{width=\maxwidth,height=\maxheight,keepaspectratio}

\usepackage{color}
\usepackage{fancyvrb}

\DefineVerbatimEnvironment{Highlighting}{Verbatim}{commandchars=\\\{\}}
\newenvironment{Shaded}{}{}

\newcommand{\FunctionTok}[1]{\textcolor[rgb]{0.02,0.16,0.49}{#1}}

\newcommand{\NormalTok}[1]{#1}
\newcommand{\OperatorTok}[1]{\textcolor[rgb]{0.40,0.40,0.40}{#1}}

\makeatletter
\@ifpackageloaded{subcaption}{}{\usepackage{subcaption}}
\@ifpackageloaded{caption}{}{\usepackage{caption}}
\AtBeginDocument{%

}
\AtBeginDocument{%

}
\newcounter{pandoccrossref@subfigures@footnote@counter}
{\end{figure}%
\addtocounter{footnote}{-\value{pandoccrossref@subfigures@footnote@counter}}
\@for\f:=\global@pandoccrossref@subfigures@footnotes\do{\stepcounter{footnote}\footnotetext{\f}}%
\gdef\global@pandoccrossref@subfigures@footnotes{}}
\@ifpackageloaded{float}{}{\usepackage{float}}
\floatstyle{ruled}
\@ifundefined{c@chapter}{\newfloat{codelisting}{h}{lop}}{\newfloat{codelisting}{h}{lop}[chapter]}
\floatname{codelisting}{Listing}

\makeatother

\begin{document}

{\centering
{\LARGE \textbf{Equilibrium Selection and Closure in DSGE Solution
Methods} \par}
\vspace{1em}
{\normalsize Mitsuhiro Okano \par}
{\normalsize Faculty of Economics, Osaka Gakuin University \par}
{\normalsize \texttt{okano@ogu.ac.jp} \par}
\vspace{1.75em}
}

\section*{Abstract}\label{abstract}
\addcontentsline{toc}{section}{Abstract}

Standard DSGE solution practice closes forward-looking equilibrium
problems by combining model specification with criteria that select one
operational equilibrium from multiple candidate paths, yet this closure
step typically remains implicit. This paper formalizes that closure step
as a substantive operation in standard DSGE solution practice. By
separating the economic specification (\(S\)), the self-referential
operator (\(T\)), and the equilibrium selection rule (\(\Pi\)), the
\((S, T, \Pi)\) triad provides a descriptive semantic grammar for
comparing closure choices across familiar solution environments. The
framework clarifies what standard solver diagnostics actually report,
captures the regime-iteration logic of piecewise-linear methods like
OccBin, explains why independent linear solvers often reproduce the same
selected equilibrium, and distinguishes within-model refinements from
specification-augmenting closure changes under indeterminacy.

\medskip

\textbf{Keywords:} DSGE solution methods, equilibrium closure, linear
rational expectations, indeterminacy

\smallskip

\textbf{JEL Classification:} C61, C63, E17, E52

\section{Introduction}\label{sec:intro}

Standard DSGE solution practice does more than translate model equations
into numerical output: it closes forward-looking equilibrium problems by
combining economic structure with criteria that select one operational
path from a larger candidate set. In standard linear rational
expectations practice, and in piecewise-linear extensions built on it,
this closure step enters through stability, boundary, or
regime-consistency conditions, yet it typically remains embedded in the
way solution procedures are presented.\footnote{See
  \citet{FernandezVillaverdeRubioRamirezSchorfheide2016} for a
  comprehensive survey of solution and estimation methods,
  \citet{Gali2015} for the New Keynesian framework, and
  \citet{Woodford2003} for theoretical foundations.} Making that step
explicit helps clarify what is being held fixed, what is being imposed,
and why neighboring solution methods can sometimes agree and sometimes
diverge.

Solver diagnostics such as ``unique stable solution,''
``indeterminacy,'' or ``no stable solution'' are one visible trace of
that closure logic, but they are not the whole point. Treating closure
as an explicit object lets the same language organize comparisons across
solution methods, separate refinements within a fixed model from changes
in the equilibrium problem itself, and clarify why alternative closure
choices may matter for downstream economic interpretation.

This paper makes explicit a distinction already operative in the DSGE
solution methods studied here: the distinction between specifying a
forward-looking equilibrium problem and closing it.

We organize this closure step around three components: the specification
(\(S\)), which encodes the constraints and optimality conditions
defining admissible outcomes; the self-referential operator (\(T\)),
which encodes how expectations about future outcomes feed back into
current equilibrium conditions; and the selection rule (\(Π\)), which
provides the criterion for choosing one equilibrium from the set
\(\mathrm{Fix}(T)\). In the forward-looking linear rational expectations
applications studied below, determinacy refers to cases in which the
relevant selector returns a unique admissible equilibrium, not to cases
in which selection disappears as a substantive step. More broadly, the
same language also helps distinguish refinements within a fixed closure
problem from changes that alter the closure conditions themselves.

These components can be introduced in the order in which solution
practice usually encounters them: economists first specify economic
structure (\(S\)), then represent how expectations about future outcomes
enter current equilibrium conditions through an operator \(T\), and
finally impose selection criteria (\(Π\)) when solving the model. This
ordering reflects how modern solution procedures actually operate and
clarifies why multiplicity, refinement, and closure are structural
features of forward-looking equilibrium analysis rather than anomalies.

The framework isolates a recurrent step in DSGE solution practice: the
closure operation that turns a forward-looking equilibrium problem into
an operational object for analysis. In the sections that follow, we use
that perspective in three ways. First, it clarifies what Blanchard--Kahn
conditions do in standard linear practice: they implement a particular
closure criterion rather than offering a general theory of equilibrium
existence \citep{BlanchardKahn1980}. Second, it helps explain why
different solution procedures, such as QZ-based solvers and OccBin
\citep{GuerrieriIacoviello2015}, can be interpreted as different
implementations of related closure logic. Third, it makes visible an
important distinction within indeterminacy analysis between within-model
refinements, such as variance-based restrictions, and alternative
closure conditions, such as fiscal anchoring after specification
augmentation.

Despite its practical importance, there is no single widely adopted
framework for isolating the closure step from the economic specification
across these neighboring literatures. This paper offers one such
language. Its contribution is comparative and diagnostic rather than
merely taxonomic: it develops an axiomatic decomposition of the closure
step, uses the \((S, T, Π)\) structure to compare algorithms such as
QZ-based solvers and OccBin, and distinguishes within-model refinements
from specification-augmenting closure changes under indeterminacy.

The remainder of this paper proceeds as follows.
Section~\ref{sec:literature} reviews related literature and positions
our contribution. Section~\ref{sec:axioms} presents the axiomatic
framework, defining DSGE systems via the \((S, T, Π)\) triad and
clarifying how determinate and indeterminate cases fit within the same
language. Section~\ref{sec:bk} formalizes Blanchard--Kahn conditions as
a selection operator \(Π_{BK}\) and provides a characterization theorem.
Section~\ref{sec:unified} compares solution methods, including QZ
decomposition, time iteration, and OccBin, within the \((S, T, Π)\)
framework. Section~\ref{sec:nk} shows that policy parameters control
eigenvalue configuration, thereby determining whether \(Π_{BK}\)
succeeds, revealing determinacy conditions as compatibility between
\(T\) and \(Π_{BK}\). Section~\ref{sec:alternative} examines how
indeterminacy invites either further refinement within a fixed model or
a re-posed closure problem after specification augmentation.
Section~\ref{sec:zlb} extends the framework to occasionally binding
constraints, showing how OccBin implements composite selection through
iterated application of \(Π_{BK}\) across regimes.
Section~\ref{sec:discussion} positions existing literature within the
\((S, T, Π)\) framework and discusses the gains from making equilibrium
closure explicit. Section~\ref{sec:conclusion} concludes. Appendix A
reports the OccBin cross-solver comparison discussed below.

\section{Related Literature}\label{sec:literature}

Several neighboring literatures describe the closure step in DSGE
analysis through different vocabularies. We focus on the strands most
directly connected to that issue: rational expectations foundations,
indeterminacy and sunspot equilibria, learning and expectation
formation, fiscal theory of the price level, regime-switching methods,
and alternative solution approaches.

The foundational work on solving rational expectations models
establishes the core solution techniques still in use today.
\citet{BlanchardKahn1980} provide conditions for saddle-path stability
in linear systems; \citet{Sims2002} develops the gensys algorithm using
QZ decomposition; \citet{Klein2000} offers an alternative
implementation. These methods implement explicit stability-based
admissibility criteria, but they do not typically separate the
equilibrium specification and the closure rule as distinct conceptual
objects. Our framework supplies a language for making that separation
precise.

A large literature documents that rational expectations models can admit
multiple equilibria. \citet{Farmer1999} shows how self-fulfilling
beliefs can generate indeterminacy;
\citet{BenhabibSchmittGroheUribe2001} characterize conditions under
which Taylor rules lead to multiple equilibria;
\citet{LubikSchorfheide2003} analyze the dimension of the sunspot
equilibrium set. This literature establishes equilibrium multiplicity as
a recurring feature of forward-looking DSGE analysis.

Several approaches explicitly address equilibrium selection through
learning dynamics or minimal state variable principles.
\citet{McCallum1983} proposes selecting equilibria that minimize state
variables; \citet{EvansHonkapohja2001} develop expectational stability
(E-stability) as a selection criterion; \citet{BranchMcGough2010} study
dynamic predictor choice in a New Keynesian learning model. These
contributions recognize selection as a substantive economic question but
tie it to specific behavioral microfoundations. Our framework abstracts
selection as a formal operation, enabling comparison across different
economic assumptions.

The fiscal theory of the price level
\citep{Leeper1991, Woodford1994, Cochrane2001, Cochrane2023FTPL} shows
how fiscal policy can determine equilibrium when monetary policy fails
to do so. Under passive monetary/active fiscal regimes, the government
budget constraint pins down the price level. Rather than treating this
as a same-system selector, our framework treats fiscal anchoring as an
alternative closure obtained after specification augmentation, making
explicit how fiscal conditions change the equilibrium problem itself.

Methods for handling regime-switching and occasionally binding
constraints extend solution techniques to nonlinear settings. OccBin
extends solution techniques to piecewise-linear systems with
occasionally binding constraints; \citet{EggertsonWoodford2003} analyze
zero lower bound equilibria; and related work such as
\citet{DavigLeeper2007} and \citet{BianchiIlut2017} studies
regime-switching or state-dependent macroeconomic environments. These
methods operationally apply solution techniques across regimes but do
not formalize the compositional structure. In Section~\ref{sec:zlb} we
show that OccBin iterates regime-specific selection operators.

Additional approaches extend DSGE solution methods in various
directions. \citet{HansenSargent2008} develop robust control methods
that select worst-case equilibria under model uncertainty.
Heterogeneous-agent and heterogeneous-expectations settings
\citep{Hommes2013, KaplanMollViolante2018} raise related questions about
how admissibility and closure are represented once the surrounding model
environment becomes richer.

Within this background, our contribution is to make the closure step
visible in the solver practices studied here. While existing work
recognizes equilibrium multiplicity and employs various techniques to
resolve it, closure criteria often remain embedded in computational
methods or behavioral assumptions rather than formalized as distinct
operations. The sections that follow therefore move from the descriptive
decomposition itself (Section~\ref{sec:axioms}), to the standard BK
closure criterion (Section~\ref{sec:bk}), to algorithmic implementations
of that closure logic (Section~\ref{sec:unified}), and finally to worked
demonstrations in standard and regime-switching environments
(Section~\ref{sec:nk} through Section~\ref{sec:zlb}).

\section{\texorpdfstring{Axiomatic Framework: The \((S, T, Π)\)
Triad}{Axiomatic Framework: The (S, T, Π) Triad}}\label{sec:axioms}

We now formalize the descriptive decomposition for the class of
forward-looking equilibrium problems studied in DSGE solution practice.
The axioms isolate the pieces needed for the diagnostic distinction
pursued here. They are stated abstractly enough to describe neighboring
settings, while the paper's formal results focus on linear rational
expectations systems and closely related extensions.

\subsection{Basic Setup}\label{sec:axioms:setup}

Let \(X\) be a space of candidate outcome sequences, with linear
rational expectations systems as the main focus below.

\nameddefinition{def:outcome-space}{Outcome Space}

Let \(X\) be a Banach space of adapted stochastic processes
\(\{x_t\}_{t=0}^∞\).\footnote{The Banach space framework aligns with
  standard treatments \citep{LjungqvistSargent2018, StokeyLucas1989} but
  completeness is not invoked in our proofs. Theorem
  \ref{thm:bk-selection} relies on linear algebra (QZ decomposition) and
  perturbation theory, not fixed-point theorems.} The space \(X\) must
support a topology under which the operators \(T\) and \(Π\) are
well-defined and continuous where needed.

Our primary focus is finite-dimensional systems in which bounded adapted
subclasses such as \(\ell^∞(ℝ^n)\) play a central role for the analysis
of stable or selected solutions. This encompasses linear RE systems
where \((A_0, A_1) ∈ ℝ^{n×n} × ℝ^{n×n}\) defines the system, and
nonlinear systems with occasionally binding constraints where state
vectors remain finite-dimensional. The same axioms apply to
infinite-dimensional specifications where \(X\) may include function
spaces such as \(L^2(S, \mathcal{F}, μ; ℝ^n)\) for state-contingent
plans, or spaces of probability measures \(\mathcal{M}\) for
heterogeneous agent models where distributions are state variables.

For the linear systems analyzed in Section~\ref{sec:bk} through
Section~\ref{sec:nk}, the finite-dimensional matrix structure
\((A_0, A_1)\) is sufficient for the formal results.

\subsection{The Four Axioms}\label{sec:axioms:four}

We now state the minimal axioms for the framework. They isolate the
formal structure needed to distinguish specification, self-reference,
and selection in the forward-looking equilibrium problems studied in
this paper.\footnote{The decomposition into
  Specification-Operator-Selection mirrors the tripartite structure of
  formal semantics in computer science: syntax (what programs are
  written), semantics (what meaning or computations they encode), and
  evaluation (which outcomes result). This analogy is conceptual rather
  than definitional; our contribution is the explicit organization of
  DSGE around these three components, enabling formal analysis of
  selection rules.}

\namedaxiom{ax:specification}{Specification}

There exists a specification \(S ⊆ X\) encoding admissible outcomes.
Formally, \(S\) is a (typically nonlinear) constraint set defined by:

\begin{equation}\protect\phantomsection\label{eq:specification}{
S = \{x ∈ X : F(x) = 0\}
}\end{equation}

where \(F: X → ℝ^m\) encodes constraints (resource feasibility, budget
constraints, aggregation conditions), optimality conditions (first-order
conditions from agent optimization, if present), and equilibrium
conditions (market clearing, policy rules, institutional constraints).

This formalizes ``the model equations'' but does not yet impose
consistency between agents' beliefs and realized outcomes. It defines
what is admissible, not what is equilibrium. The constraint-set
formulation \eqref{eq:specification} is standard in dynamic optimization
and equilibrium theory \citep{LjungqvistSargent2018}, encompassing
equality constraints, complementarity restrictions, and both finite- and
infinite-dimensional systems.

\namedaxiom{ax:self-reference}{Self-Reference}

There exists an operator \(T: X → X\) such that equilibria are fixed
points: \(x^* ∈ X\) is an equilibrium if and only if \(x^* = T(x^*)\).
The operator \(T\) embeds forward-looking expectations into current
outcomes. The fixed-point formulation \(x = T(x)\) is standard in
dynamic economic systems \citep{LjungqvistSargent2018, Woodford2003}.

Formal structure: Let \(\mathcal{E}_t: X → ℝ^n\) denote the expectation
operator (conditioning on information at \(t\)), representing the
conditional expectation \(E[· | \mathcal{F}_t]\) in stochastic settings.
Then \(T\) satisfies:\footnote{We distinguish three notational forms:
  \(\mathcal{E}_t\) denotes the expectation operator as a formal object
  (Axiom \ref{ax:self-reference}, Definition \ref{def:ree-property});
  \(E_t[\cdot]\) denotes conditional expectations in model equations (NK
  specification, matrix forms); \(\mathbb{E}[\cdot]\) denotes
  expectations in functional contexts (variance minimization,
  optimization). All refer to the same concept; notation reflects
  contextual usage.}

\begin{equation}\protect\phantomsection\label{eq:operator}{
(Tx)_t = G(x_t, \mathcal{E}_t[x_{t+1}], ε_t)
}\end{equation}

for some function \(G\) and exogenous shocks \(ε_t\). Equation
\eqref{eq:operator} maps a candidate outcome sequence \(x\) to a new
sequence where period-\(t\) outcomes depend on expected future outcomes
\(\mathcal{E}_t[x_{t+1}]\).

While the axiom is stated abstractly to accommodate time-varying
dynamics where \(T\) may depend on time or system state, our formal
results focus on time-invariant systems. This axiom formalizes
expectations feedback rather than committing the framework in advance to
a single expectation-formation mechanism; rational expectations is
introduced explicitly in Definition \ref{def:ree-property}. Agents form
expectations about the process \(x\), and those expectations feed back
into determining \(x\). This creates circularity: to know \(x\), we need
to know what agents expect about \(x\), but expectations depend on \(x\)
itself.\footnote{In computational practice, \(T\) is typically
  constructed as a composition of transformations: parsing the
  specification, linearizing around steady state, extracting canonical
  form, etc. This pipeline can be conceptualized as
  \(T = T_1 \circ T_0\) where \(T_0\) performs symbolic transformations
  (equation manipulation, substitution) and \(T_1\) imposes dynamic
  structure (time indices, expectations operators, state-space
  representation). Mathematically, we treat the composed result as a
  single operator \(T: X → X\) for notational clarity.
  Section~\ref{sec:unified} and Appendix A discuss this compositional
  structure in implementation contexts.}

\namedaxiom{ax:multiplicity}{Multiplicity}

For many economically relevant specifications, the fixed-point set
\(\mathrm{Fix}(T) ≡ \{x ∈ X : x = T(x)\}\) is non-singleton. That is,
the self-referential structure encoded by \(T\) can generate multiple
solutions to \(x = T(x)\). It may contain explosive candidate paths
where \(\|x_t\| → ∞\) (violating transversality conditions), sunspot
equilibria driven by non-fundamental randomness, and multiple stable
solutions that remain bounded when stability conditions are weak.
Degenerate cases in which \(\mathrm{Fix}(T) = \{x^*\}\) can also be
accommodated, but they are not the primary forward-looking cases. This
framework highlights how multiplicity and admissibility interact across
different environments.

This captures the fundamental indeterminacy potential of forward-looking
systems with expectations feedback. Without additional restrictions,
such models can admit a broad set of candidate equilibria. This is not a
pathology but rather the key reason a selection operator becomes
methodologically important. In the forward-looking settings, determinacy
instead refers to the case in which the relevant admissibility criterion
selects a unique equilibrium from that broader candidate set.

\namedaxiom{ax:selection}{Selection}

There exists a family \(\mathcal{D}_Π \subseteq 2^X\) of equilibrium
sets on which the relevant selection rule is defined, together with a
partial selection operator \(Π: \mathcal{D}_Π \rightharpoonup X\), such
that whenever \(\mathrm{Fix}(T) \in \mathcal{D}_Π\),

\begin{equation}\protect\phantomsection\label{eq:selection}{
x^* = Π(\mathrm{Fix}(T))
}\end{equation}

The operator \(Π\) maps a fixed-point set in its domain to a single
selected element. Formally, \(Π\) encodes stability criteria
(transversality conditions, bounded-growth restrictions, no-Ponzi-game
constraints), boundary conditions (initial values for predetermined
variables, terminal conditions), and refinement principles (among
multiple stable solutions, select via optimality, policy objectives, or
other criteria). When \(\mathrm{Fix}(T) \notin \mathcal{D}_Π\), the
selector is not well-defined for that problem under the maintained
criterion.

This axiom formalizes the selection problem. Given multiple fixed
points, which one should we use for analysis, forecasting, or policy
evaluation? Different selectors \(Π\) correspond to different economic
assumptions or solution concepts. When \(\mathrm{Fix}(T)\) is singleton,
the selector is trivial; in the forward-looking applications studied
here, however, its substantive role is to isolate a bounded, stable, or
otherwise refined element from a larger fixed-point set.

Boundedness therefore enters naturally at the level of selection or
admissibility, not necessarily at the level of the ambient outcome space
itself. This is why explosive paths may belong to \(\mathrm{Fix}(T)\)
even though economically relevant analysis often focuses on bounded or
stable subclasses.

Finally, we note that while not a separate axiom, the operator \(T\) in
macroeconomic applications almost always encodes temporal
structure---period-\(t\) outcomes depend on expectations about period
\(t+1\). Static simultaneous-equation systems can be viewed as
degenerate cases where \(T\) has no forward-looking component. We focus
on dynamic systems where temporal ordering is essential.

\subsection{Definition of the Framework}\label{sec:axioms:definition}

\nameddefinition{def:dsge-triad}{Descriptive DSGE System}

For the class of forward-looking equilibrium problems studied here, we
represent a DSGE system by \((X, S, T, Π)\) where:

\begin{itemize}
\tightlist
\item
  \(X\) is an outcome space (Definition \ref{def:outcome-space})
\item
  \(S ⊆ X\) is a specification (Axiom \ref{ax:specification})
\item
  \(T: X → X\) is a self-referential operator (Axiom
  \ref{ax:self-reference})
\item
  \(Π: \mathcal{D}_Π \rightharpoonup X\) is a partial selection operator
  on a family \(\mathcal{D}_Π \subseteq 2^X\) (Axiom \ref{ax:selection})
\end{itemize}

with Axiom \ref{ax:multiplicity} highlighting the economically relevant
possibility that \(\mathrm{Fix}(T)\) may be non-singleton.

When the outcome space \(X\) is clear from context (as is typical in
applications), we write \((S, T, Π)\) and refer to this as the DSGE
triad. This notation emphasizes the descriptive core needed for the
present paper: the specification (\(S\)) that encodes constraints, the
self-referential operator (\(T\)) that embeds forward-looking behavior,
and the selection rule (\(Π\)) that resolves equilibrium multiplicity.
The space \(X\) is a technical prerequisite but does not alter the basic
operation of fixed-point selection from a self-referential system.

Economic content---preferences, technologies, optimization---enters
through how we construct \(S\) and \(T\). The purpose of the triad is to
provide a disciplined descriptive decomposition for the present
argument.

\subsection{Formal Necessity of the
Distinction}\label{sec:axioms:minimality}

The diagnostic distinction pursued here requires all four components.
Removing any one erases either the self-referential structure, the
specification layer, or the language needed to distinguish determinate
from indeterminate uses of the model.

\begin{proposition}[Formal Necessity for Diagnostic Separation]\label{prop:formal-necessity}
The components $(X, S, T, \Pi)$ are necessary for the following distinctions:

\begin{enumerate}
\item[(i)] \textit{Without Specification $S$: If $S = X$ (no constraints), then any outcome $x \in X$ is admissible. The model imposes no restrictions, yielding no economic content. The specification becomes vacuous.}
\item[(ii)] \textit{Without Self-Reference $T$: If there is no operator $T$ (or $T$ is the identity), there is no forward-looking component. Outcomes are determined by $S$ alone, reducing to a static system or a recursion without expectational feedback. This eliminates the "dynamic stochastic general equilibrium" character: no rational expectations and no equilibrium indeterminacy arising from self-reference.}
\item[(iii)] \textit{Without Multiplicity (Axiom \ref{ax:multiplicity}): If one rules out the possibility that $\mathrm{Fix}(T)$ can be non-singleton, then the framework collapses to determinate special cases only. In such cases $\Pi$ becomes the identity map, and the language needed to describe indeterminacy, sunspots, or explosive candidate paths disappears. The resulting system cannot distinguish the determinate cases of interest from the indeterminate ones that motivate equilibrium selection.}
\item[(iv)] \textit{Without Selection $\Pi$: If $\mathrm{Fix}(T)$ is non-singleton but no selection rule is specified, the framework does not by itself determine a unique object for simulation, forecasting, or policy analysis. One may still study the resulting equilibrium set directly, but without the language of $\Pi$ the framework cannot state the difference between a case where no extra restriction is needed and a case where boundedness, transversality, inference, or another refinement is doing the closing work.}
\end{enumerate}

  Thus, each component $(S, T, \Pi)$ plays an irreducible role in the diagnostic architecture of the framework used here. Removing any one either eliminates essential features of the class of problems under study or erases the distinction between cases where selection is trivial and cases where it is substantive.
\end{proposition}

\subsection{Rational Expectations Equilibrium (REE) as Property, Not
Primitive}\label{sec:axioms:ree}

One common usage treats DSGE analysis as if it were synonymous with
rational expectations equilibrium. The present framework keeps those
notions distinct.

\nameddefinition{def:ree-property}{REE Property}

A fixed point \(x^* ∈ \mathrm{Fix}(T)\) satisfies the Rational
Expectations Equilibrium property if the expectation operator
\(\mathcal{E}_t\) embedded in \(T\) coincides with the conditional
expectation under the probability law governing \(x^*\).

That is, agents' subjective beliefs (encoded in \(\mathcal{E}_t\)) equal
the objective distribution of outcomes (generated by \(x^*\)). This is
the ``rational expectations'' assumption: agents know the model and form
expectations consistent with the model's implications.

The crucial distinction is that REE is a property of a fixed point, not
a defining feature of the \((S, T, Π)\) framework. The framework
accommodates REE fixed points with beliefs
\(\mathcal{E}_t = E[· | \mathcal{F}_t]\) equaling realized outcomes. It
can also be extended beyond that case to learning dynamics in which
\(\mathcal{E}_t\) evolves over time as agents update beliefs and \(T\)
may depend on time or past forecast errors
\citep{EvansHonkapohja2001, BranchMcGough2010}. Heterogeneous
expectations can likewise be represented by allowing different
forecasting rules to enter \(\mathcal{E}_t\) or its law of motion. The
formal results and main applications of this paper, however, remain
focused on linear rational expectations settings and nearby
extensions.\footnote{Modern DSGE extensions incorporate learning,
  heterogeneous expectations, inattention, and liquidity constraints
  \citep{EvansHonkapohja2001, Hommes2013, Gabaix2020, KaplanMollViolante2018}
  through modifications to \(\mathcal{E}_t\), \(S\), or the induced
  operator \(T\). Developing those extensions formally is beyond the
  present paper's scope.}

\section{Blanchard--Kahn as Selection Operator}\label{sec:bk}

We now turn to the closure criterion that sits at the center of standard
linear rational expectations solver practice: the Blanchard--Kahn (BK)
conditions. Rather than treating BK only as a determinacy diagnostic, we
show how it can be represented here as the operative closure rule
applied to the fixed-point problem.

\subsection{\texorpdfstring{Linear Rational Expectations Setting and the
\(Π_{BK}\)
Operator}{Linear Rational Expectations Setting and the Π\_\{BK\} Operator}}\label{sec:bk:linear}

Consider a linear rational expectations system in canonical form:

\begin{equation}\protect\phantomsection\label{eq:canonical}{
A_0 z_t = A_1 E_t[z_{t+1}] + B ε_{t}
}\end{equation}

where \(z_t ∈ ℝ^n\) is the vector of endogenous variables, \(ε_{t}\) is
a vector of exogenous shocks, and \(A_0, A_1 ∈ ℝ^{n×n}\) and
\(B ∈ ℝ^{n×k}\) are coefficient matrices. We partition
\(z_t = [s_t; j_t]\) where \(s_t ∈ ℝ^{n_s}\) are predetermined (state)
variables (given at time t) and \(j_t ∈ ℝ^{n_j}\) are jump
(non-predetermined) variables (chosen at time t), with
\(n_s + n_j = n\). The predetermined/jump distinction is fundamental to
selection. Predetermined variables have known initial conditions; jump
variables are free at \(t=0\) and must be pinned down by stability
requirements.

For a given specification, the fixed-point set is generally larger than
the economically admissible solution set. In linear rational
expectations systems, candidate fixed points can include saddle-path
stable solutions, mathematically valid but economically inadmissible
explosive paths, and, under indeterminacy, sunspot equilibria driven by
non-fundamental shocks. This is the sense in which closure matters: the
model equations alone do not determine which candidate path is retained
for analysis.

\nameddefinition{def:bk-selector}{Blanchard--Kahn Selection Operator}

On the class of linear RE systems for which the BK admissibility
criterion selects a unique saddle-path stable solution, \(Π_{BK}\) is
defined by:

\begin{enumerate}
\def\labelenumi{\arabic{enumi}.}
\tightlist
\item
  Set all martingale/explosive components to zero
\item
  Set initial conditions for predetermined variables \(s_0\)
\item
  Compute jump variables \(j_0\) to satisfy saddle-path stability
\end{enumerate}

Formally, \(Π_{BK}\) projects onto the stable eigenspace of the system.
In the determinate linear cases studied below, this selector is
well-defined and returns the unique saddle-path stable equilibrium;
outside that domain, BK diagnostics report why the selector is not
well-defined.

\subsection{QZ Decomposition and Spectral Filtering}\label{sec:bk:qz}

To implement \(Π_{BK}\) computationally, we use the Generalized Schur
(QZ) decomposition of the matrix pencil \((A_0, A_1)\), a standard
approach implemented in modern DSGE solvers \citep{Klein2000}. There
exist unitary matrices \(Q\), \(Z\) such that \(Q A_0 Z = S\) (upper
triangular) and \(Q A_1 Z = T\) (upper triangular). Generalized
eigenvalues are \(\displaystyle λ_i = \frac{T_{ii}}{S_{ii}}\) (defined
when \(S_{ii}≠0\)). Using \texttt{ordqz} to reorder so that eigenvalues
with \(|λ| < 1\) appear first partitions the system into a stable block
(\(|λ| < 1\), first \(n_s\) eigenvalues) and an unstable block
(\(|λ| ≥ 1\), remaining \(n_j\) eigenvalues).

The \(Π_{BK}\) selection rule succeeds if and only if: (1)
count\((|λ| < 1) = n_s\) (dimension match); (2) The stable subspace is
spanned by \(Z\) correctly partitioned; (3) \(Z_{11}\) is invertible
(technical condition). When these hold, the policy function is
\(j_t = Z_{21} Z_{11}^{-1} s_t = P s_t\) where \(P\) is the policy
matrix.

\subsection{BK-as-Selection Theorem}\label{sec:bk:theorem}

The theorem below gives a domain-restricted characterization of when the
Blanchard--Kahn stability criterion defines a well-defined selection
operator on \(\mathrm{Fix}(T)\) and when that selector varies
continuously with model parameters. Assumption (A1) governs regularity
of the linear representation itself and is best read as a prerequisite
on the underlying system representation; assumptions (A2)--(A4) govern
whether \(\Pi_{BK}\) is well-defined as a stability selector.

\begin{theorem}[BK Selection Operator --- Characterization]\label{thm:bk-selection}
  Let $(A_0, A_1)$ define a linear RE system with state/jump partition $(n_s, n_j)$. On the class of systems satisfying the following standard conditions, $\Pi_{BK}$ constitutes a well-defined, generically continuous selection operator that recovers the unique saddle-path stable solution in the determinate linear case. Assume:

\begin{enumerate}
\item[(A1)] \textit{Regular representation condition: $\det(A_0 - \lambda A_1)$ is not identically zero.}
\item[(A2)] \textit{Spectral alignment condition: $\#\{|\lambda| < 1\} = n_s$.}
\item[(A3)] \textit{Rank condition with spectral gap: The stable subspace (columns of $Z$ corresponding to $|\lambda| < 1$) has full rank $n_s$ when restricted to the predetermined block (ensuring $Z_{11}$ invertibility), and stable eigenvalues are separated from unstable ones by a nonzero gap (no clustering on the unit circle).}
\item[(A4)] \textit{No unit roots: No generalized eigenvalues satisfy $|\lambda| = 1$ exactly (or they are handled via cointegration/preprocessing).}
\end{enumerate}

  Then:

\begin{enumerate}
\item \textit{Well-definedness of $\Pi_{BK}$: There exists a unique policy matrix $P$ such that $j_t = P s_t$ is the $\Pi_{BK}$-selected solution.}
\item \textit{Continuity (generic): $\Pi_{BK}(A_0, A_1)$ varies continuously with $(A_0, A_1)$ on open sets where the spectral gap and rank conditions hold; discontinuities arise only at bifurcation points where eigenvalues collide or cross $|\lambda| = 1$.}
\item \textit{Determinacy: In the linear determinate case singled out by (A2), the selected equilibrium is the unique saddle-path stable solution.}
\end{enumerate}
\end{theorem}

\proofheading{Proof sketch}

Part 1 (Structure of Fix(T)): Under indeterminacy
(\(\#\{|λ| < 1\} > n_s\)), the fixed-point set admits an affine
structure \citep{LubikSchorfheide2003}:

\begin{equation}\protect\phantomsection\label{eq:sunspot}{z_t = z_t^f + W\xi_t}\end{equation}

where \(z_t^f\) is the fundamental solution and \(W\) spans the excess
stable eigenspace. The spectral decomposition \citep{Klein2000}
decouples the system so that eigenvalues \(|\lambda_i| > 1\) admit
families of solutions parameterized by arbitrary sunspot noise
\(\xi_{it}\) satisfying \(\xi_{it} = \lambda_i^{-1} E_t[\xi_{i,t+1}]\).
Selection is necessary because neither \(S\) nor \(T\) pins down
\(\xi_t\).

Part 2 (Well-definedness of \(\Pi_{BK}\)): \(\Pi_{BK}\) selects the
fundamental solution by setting \(\xi_t = 0\). Here (A1) ensures that
the linear representation itself admits a valid spectral decomposition,
while (A2)-(A3) ensure that the stability selector can be applied
without ambiguity. Under (A1)-(A3):

Spectral decomposition exists by (A1) \citep{Klein2000}, Stable block
has dimension \(n_s\) by (A2), enabling correct state/jump partition,
and \(Z_{11}\) is invertible by (A3), so \(P = Z_{21} Z_{11}^{-1}\) is
well-defined \citep[Proposition 1]{Klein2000}

Part 3 (Uniqueness and Continuity): Under the spectral alignment
condition (A2), the excess stable directions responsible for
indeterminacy are absent, so the stable-subspace restriction and initial
conditions jointly pin down a unique admissible solution. Continuity
follows from perturbation theory \citep{Sims2002}: the map
\((A_0, A_1) \mapsto P\) is continuous wherever the spectral gap (A3)
persists. Discontinuities arise when eigenvalues cross \(|\lambda| = 1\)
(bifurcation), violating the conditions under which the selector is
well-defined. \proofend

Interpretation: \(\Pi_{BK}\) is a stability filter selecting the unique
non-explosive element from the broader set \(\mathrm{Fix}(T)\).
Bifurcations represent selector failure, not equilibrium non-existence.

\subsection{Reinterpreting Solver Diagnostics}\label{sec:bk:diagnostics}

Standard DSGE solvers (Dynare, gensys) report one of three outcomes.
``Unique stable solution'' appears when count(\(|λ| < 1) = n_s\),
indicating \(Π_{BK}\) is well-defined. ``Indeterminacy'' appears when
count(\(|λ| < 1) > n_s\), indicating \(\mathrm{Fix}(T)\) contains a
subspace of stable solutions and \(Π_{BK}\) is not well-defined because
the stability criterion does not select uniquely. ``No stable solution''
appears when count(\(|λ| < 1) < n_s\), indicating \(Π_{BK}\) is not
well-defined because no solution satisfies both stability and initial
conditions.

Key reinterpretation (Theorem \ref{thm:bk-selection}) reveals these are
statements about \(Π_{BK}\), not about the specification \(S\) or
operator \(T\) per se. The two failure cases are distinct: in one, the
admissibility criterion is too coarse to isolate a unique element; in
the other, it is too restrictive to admit any element at all. BK
conditions are therefore selection criteria, not existence/uniqueness
theorems about equilibria in general. When \(Π_{BK}\)
fails---particularly under indeterminacy---additional responses may be
considered. Section~\ref{sec:alternative} discusses one within-model
refinement (minimal-variance \(Π_{MV}\)) and one augmented fiscal
closure, illustrating that standard stability selection does not exhaust
the ways one might restrict or reclose the equilibrium problem.

\subsection{Distinction: Existence vs Selection}\label{sec:bk:existence}

Classical existence theorems in economic theory establish non-emptiness
of equilibrium sets under regularity conditions. Arrow--Debreu
\citep{ArrowDebreu1954}, for instance, proves existence of Walrasian
equilibria given continuity, compactness, and convexity. Analogously,
one might ask whether \(\mathrm{Fix}(T)\) is non-empty in rational
expectations models. BK conditions, however, do not address existence of
\(\mathrm{Fix}(T)\)---they address which element of \(\mathrm{Fix}(T)\)
to select. For example, the forward-looking equation
\(y_t = E_t[y_{t+1}] + ε_{t}\) admits infinitely many fixed points
(unique forward solution, infinite backward solutions).
BK/transversality selects the forward solution by ruling out explosive
paths. This is selection, not existence.

\section{\texorpdfstring{Comparing Solution Methods Through
\((S, T, Π)\)}{Comparing Solution Methods Through (S, T, Π)}}\label{sec:unified}

DSGE solution methods---linear QZ decomposition, time iteration, and
piecewise-linear regime-switching (OccBin)---can be compared through the
\((S, T, Π)\) framework. They do not all carry the same formal
guarantees, but they can still be described in a common language that
separates model specification, self-reference, and admissibility.

\subsection{Two Forms of Comparison}\label{sec:unified:forms}

The \((S, T, Π)\) framework connects DSGE solution methods in two
distinct senses.

Implementation equivalence applies to linear RE systems where different
solvers produce identical results because they implement the same
algorithm. QZ, gensys, and Dynare agree numerically because they execute
the same \((S, T, Π_{BK})\) via deterministic operations (linearization,
eigendecomposition, stable projection). This helps explain the
reproducibility discussed in Section~\ref{sec:unified:identical}.

Descriptive decomposition applies to the methods discussed here. Each of
them can be described in terms of constructing \(T\), characterizing
\(\mathrm{Fix}(T)\), and applying \(Π\). Outside linear RE systems,
these steps admit multiple implementations; the framework provides a
common language, not a universal algorithm
(Section~\ref{sec:unified:occbin} and
Section~\ref{sec:unified:pattern}).

What follows demonstrates both forms. We show implementation equivalence
for linear solvers (Section~\ref{sec:unified:qz} and
Section~\ref{sec:unified:identical}), then descriptive comparison for
piecewise-linear methods (Section~\ref{sec:unified:occbin}) and
neighboring solution approaches (Section~\ref{sec:unified:pattern}).

\subsection{Linear Selection via QZ Decomposition}\label{sec:unified:qz}

In linear rational expectations systems, the QZ decomposition
\citep{Klein2000, Sims2002} provides the standard implementation of
\(Π_{BK}\). Implemented in gensys and Dynare, this method directly
realizes the selection operator defined in Section~\ref{sec:bk} by
restricting the solution to the stable subspace.

The operator \(T\) is encoded implicitly in the matrix pencil
\((A_0, A_1)\) satisfying \(A_0 z_t = A_1 E_t[z_{t+1}] + B ε_{t}\). The
QZ decomposition solves for the policy function \(P\) such that
\(j_t = P s_t\) and \(s_{t+1} = R s_t + Q ε_{t+1}\) via eigenvalue and
eigenvector analysis, thereby finding elements of \(\mathrm{Fix}(T)\)
satisfying the linear difference equation.

The QZ method implements \(Π_{BK}\) by first decomposing \((A_0, A_1)\)
to extract generalized eigenvalues \(λ_i\). By reordering the
eigenvalues so that \(|λ| < 1\) appear first (spectral filtering) and
counting the stable eigenvalues (which must equal \(n_s\)), it restricts
the solution via \(P = Z_{21} Z_{11}^{-1}\) to the stable manifold. This
realizes the \(Π_{BK}\) operator from Definition \ref{def:bk-selector}
by selecting the admissible solution associated with that stable
manifold when the BK alignment conditions hold.

Geometrically, the operator \(Π_{BK}\) can be read as eliminating
unstable directions and restricting attention to the stable manifold in
the state-space. The resulting policy function lies entirely in the
stable subspace, satisfying both the difference equation (membership in
\(\mathrm{Fix}(T)\)) and the admissibility conditions encoded by \(Π\).

The method produces four key diagnostics implementing assumptions A1-A4
from Theorem \ref{thm:bk-selection}. Eigenvalue plots distinguish stable
(inside unit circle) from unstable roots. Dimension matching compares
count(stable) versus \(n_s\) (assumption A2). Condition numbers flag
near-singularity via cond(\(Z_{11}\)) (assumption A3). Unit root
detection flags cases where \(|λ| ≈ 1\) (assumption A4).

As noted in Section~\ref{sec:axioms}, computational implementations
naturally decompose \(T\) into a pipeline (specification parsing,
symbolic manipulation, canonical form, fixed-point computation).
Similarly, \(Π\) may involve multiple selection stages (stability
filter, refinement, variance minimization). The \((S, T, Π)\) framework
accommodates both the mathematical abstraction (single operators) and
the computational pipeline (operator compositions), linked by the common
descriptive pattern of Section~\ref{sec:unified:pattern}.

\subsection{Piecewise-Linear Systems: OccBin as Iterated
Selection}\label{sec:unified:occbin}

When constraints bind occasionally (e.g., zero lower bound on interest
rates), the operator \(T\) becomes piecewise-linear: different regimes
have different linear systems.\footnote{OccBin is a solution method for
  occasionally binding constraints via regime iteration. This subsection
  introduces the conceptual structure; Section~\ref{sec:zlb} presents
  the full formalization with implementation details (Appendix A).}

OccBin extends \(Π_{BK}\) to this setting via regime iteration by
applying \(Π_{BK}\) within each regime and then verifying regime
consistency; Section~\ref{sec:zlb} gives the fuller formalization used
in the paper. The result is a composite selector:

\[Π_{OccBin} = Π_{BK} ∘ Π_{regime}\]

where \(Π_{BK}\) performs stable manifold projection within each regime
(Section~\ref{sec:unified:qz}) and \(Π_{regime}\) enforces
regime-consistency.

For the purposes of this comparison, OccBin can be read as reusing the
same BK-style stability filter iteratively across regimes. The local
stability logic remains the same, while the specification \(S\) and
operator \(T\) become regime-dependent.

Geometrically, the solution manifold is kinked---smooth within each
regime (stable manifold), discontinuous at boundaries (constraint
binds/unbinds). OccBin finds the unique path that stays on the stable
manifold within each regime while satisfying constraint complementarity
at transitions.

This comparison shows that BK-style local stability logic extends
naturally to piecewise-linear settings when regime iteration converges.
It does not by itself establish unrestricted portability across distinct
solution architectures.

\subsection{The Semantic Pattern}\label{sec:unified:pattern}

Beyond implementation details, the solution methods discussed here share
a common descriptive pattern. First, they construct the operator \(T\)
from the specification \(S\) to encode how expectations feed back into
outcomes. For linear systems this corresponds to linearization into the
\((A_0, A_1)\) matrix pencil, while for nonlinear systems it may involve
discretization or projection. Second, they characterize candidate
equilibria satisfying \(x = T(x)\). Linear systems extract eigenvalues
via QZ decomposition, whereas nonlinear systems might rely on iteration
or collocation. Third, they apply the selection operator \(Π\) to choose
one equilibrium from \(\mathrm{Fix}(T)\). Where linear theory applies
\(Π_{BK}\) to project onto the stable manifold, other approaches may
enforce transversality, minimize a loss function, or apply fiscal
constraints.

For linear RE systems (Section~\ref{sec:unified:qz} and
Section~\ref{sec:unified:identical}), this conceptual sequence is fully
deterministic---\(S\) uniquely determines \(T\), and \(Π_{BK}\) is
uniquely defined by Theorem \ref{thm:bk-selection}. Outside this
setting, the pattern remains the same but admits implementation
flexibility through different discretization schemes or selection
criteria.

\subsection{Why Linear Solvers Produce Identical
Results}\label{sec:unified:identical}

The reproducibility puzzle presents a fundamental empirical fact about
linear DSGE solvers. Gensys (MATLAB/Julia), Dynare (preprocessor + C++),
and manual QZ implementations produce identical impulse responses up to
numerical precision (\(<10^{-8}\)). This holds across different
programming languages (MATLAB, Julia, C++, Python), different linear
algebra libraries (LAPACK, Eigen, NumPy), and different syntaxes
(Dynare's \texttt{.mod} files vs gensys's matrix input). Standard
explanations appeal to ``mathematical equivalence'' but do not make the
equivalence precise.

The solvers produce identical results because they implement the same
underlying logic. To construct \(T\) from \(S\), Dynare parses
\texttt{.mod} files and applies symbolic differentiation to generate
\((A_0, A_1, B)\); gensys takes \((A_0, A_1, B)\) directly as input; and
manual implementations derive \((A_0, A_1, B)\) directly from economic
equations. In all cases, the encoded specification \(S\) yields an
identical matrix pencil (up to row/column permutations).

To compute \(\mathrm{Fix}(T)\), all solvers use the QZ decomposition via
LAPACK routines (e.g., \texttt{dgges}) to extract generalized
eigenvalues \(\lambda_i = T_{ii}/S_{ii}\) and reorder them. Given the
deterministic nature of the QZ algorithm, identical input matrices
produce the same eigenvalues and eigenvectors.

Finally, to apply \(\Pi_{BK}\), the solvers count stable eigenvalues
\(n_{stable} = \#\{ |\lambda| < 1\}\), check the dimension match
\(n_{stable} \stackrel{?}{=} n_s\), and compute the policy matrix
\(P = Z_{21} Z_{11}^{-1}\). Because the selection operator \(\Pi_{BK}\)
is uniquely defined by Theorem \ref{thm:bk-selection}, there is only one
stable manifold projection satisfying assumptions A1-A4.

This structural explanation has an important implication. The
\((S, T, \Pi)\) framework helps organize why solver alignment is often
observed rather than merely restating that it occurs. When solvers
agree, they do so because they implement the same closure logic despite
differing implementations. To verify a new solver, compare its
constructed matrices \((A_0, A_1)\), computed eigenvalues \(\lambda_i\),
and policy matrix \(P\) against a reference solver. Alignment on all
three is strong evidence of correct implementation of
\((S, T, \Pi_{BK})\).

\subsection{\texorpdfstring{Comparison Table: Methods as \((S, T, Π)\)
Instances}{Comparison Table: Methods as (S, T, Π) Instances}}\label{sec:unified:comparison}

Table~\ref{tbl:methods} summarizes core solution methods within the
\((S, T, Π)\) framework. For literature context and key references, see
Section~\ref{sec:discussion}.

\textbf{{[}Table~\ref{tbl:methods} here{]}}

Table~\ref{tbl:methods} shows how each method in the table can be
described within the \((S, T, Π)\) vocabulary despite varying
implementations. While the operator \(T\) varies (linear matrix,
functional, composite) and the computation of \(\mathrm{Fix}(T)\)
differs (eigendecomposition, iteration, optimization), the same
three-step structure recurs across the methods discussed here. What
carries over most directly is not unrestricted selector
interchangeability, but a comparable way of identifying specification,
self-reference, and admissibility. Operational portability is strongest
within families that share the same underlying representation, whereas
other rows in the table are included at the level of descriptive
comparison only. This structural comparison supports more consistent
diagnostics and clearer extension paths to new methods.

The framework can now be illustrated with a canonical example.
Section~\ref{sec:nk} shows that the \((S, T, Π_{BK})\) decomposition
reproduces familiar determinacy regions, eigenvalue behavior, and solver
diagnostics in the standard New Keynesian model.

\section{New Keynesian Example: Determinacy and Spectral
Structure}\label{sec:nk}

The canonical three-equation New Keynesian model demonstrates that
policy parameter \(φ_π\) controls eigenvalue configuration, thereby
determining whether \(Π_{BK}\) succeeds. Standard determinacy conditions
are compatibility conditions between \(T\) (which embeds policy) and
\(Π_{BK}\) (which imposes stability).

\subsection{\texorpdfstring{The NK Specification as
\((S, T, \Pi_{BK})\)}{The NK Specification as (S, T, \textbackslash Pi\_\{BK\})}}\label{sec:nk:spec}

Following the decomposition introduced in Section~\ref{sec:axioms}, the
standard New Keynesian model can be written directly in terms of the
three components of the \((S, T, \Pi)\) triad. The NK specification
\(S\) consists of the IS curve, Phillips curve, Taylor rule, and natural
rate process \citep{Woodford2003, Gali2015}: \[
\begin{aligned}
y_t &= E_t[y_{t+1}] - (1/σ)(i_t - E_t[π_{t+1}] - r_t^n) \\
π_t &= β E_t[π_{t+1}] + κ y_t \\
i_t &= φ_π π_t + φ_y y_t \\
r_t^n &= ρ r_{t-1}^n + ε_t
\end{aligned}
\]

where \(y_t\) is the output gap, \(π_t\) inflation, \(i_t\) nominal
interest rate, and \(r_t^n\) the natural rate of interest. Substituting
the Taylor rule into the IS equation introduces forward-looking
expectations, creating the self-referential operator \(T\) encoded in
the matrix pencil \((A_0, A_1, B)\). The system has one predetermined
variable (\(r_t^n\), so \(n_s = 1\)) and two jump variables
(\(y_t, \pi_t\), so \(n_j = 2\)); \(\Pi_{BK}\) succeeds iff
count\((|\lambda| < 1) = 1\).

In this setup, the policy parameters \((φ_π, φ_y)\) alter the operator
\(T\) itself by shifting the generalized eigenvalues, which determines
whether \(\Pi_{BK}\) succeeds. The Taylor principle thus functions as a
compatibility condition between \(T\) and \(\Pi_{BK}\).

\subsection{Matrix Representation and Regularity}\label{sec:nk:matrix}

Substituting the Taylor rule into the IS curve and writing in
expectational form \(A_0 z_t = A_1 E_t[z_{t+1}] + B ε_t\) with
\(z_t = [y_t, π_t, r_t^n]^T\) yields:

\begin{equation}\protect\phantomsection\label{eq:nk-matrices}{
A_0 = \begin{bmatrix} 
1 + φ_y/σ & φ_π/σ & -1/σ \\ 
-κ & 1 & 0 \\ 
0 & 0 & 1 
\end{bmatrix}, \quad 
A_1 = \begin{bmatrix} 
1 & 1/σ & 0 \\ 
0 & β & 0 \\ 
0 & 0 & ρ 
\end{bmatrix}, \quad 
B = \begin{bmatrix} 
0 \\ 
0 \\ 
1 
\end{bmatrix}
}\end{equation}

Regularity verification confirms that \(\det(A_0) = 1 + φ_y/σ > 0\) for
all \(φ_y ≥ 0\), so \((A_0, A_1)\) is a regular pencil with well-defined
generalized eigenvalues \(λ\) solving \(\det(A_1 - λ A_0) = 0\). This
satisfies assumption (A1) in Theorem \ref{thm:bk-selection}, ensuring
\(Π_{BK}\) is well-defined.

The matrices \((A_0, A_1)\) derive mechanically from the specification
\(S\). Importantly, the policy parameters \(φ_π\) and \(φ_y\) appear
directly in these matrices. This means the operator \(T\) itself (not
merely equilibrium outcomes) shifts with policy. When the central bank
changes \(φ_π\), it does not merely adjust outcomes; it fundamentally
alters the self-referential structure \(T\). The policy parameter
\(φ_π\) appears in \(A_0\) (row 1, column 2), directly affecting the
characteristic polynomial. Changes in \(φ_π\) shift eigenvalues in the
complex plane, altering the count of stable eigenvalues and thereby the
success/failure of \(Π_{BK}\). Section~\ref{sec:nk:policy} exploits this
to show that the Taylor principle is a condition on eigenvalue count
relative to predetermined dimensions, hence a compatibility condition
between \(T\) (which depends on policy) and \(Π_{BK}\) (which imposes
stability), not an existence or uniqueness theorem about the economic
model itself.

\subsection{\texorpdfstring{Policy Parameters and \(Π_{BK}\)
Success}{Policy Parameters and Π\_\{BK\} Success}}\label{sec:nk:policy}

Does changing policy parameters \((φ_π, φ_y)\) alter the success of
\(Π_{BK}\) as predicted by the framework? Policy parameters enter the
matrix pencil \((A_0, A_1)\), thereby shifting generalized eigenvalues
\(λ\) solving \(\det(A_1 - λ A_0) = 0\). The operator \(T\) (whose fixed
points satisfy these eigenvalues) changes continuously with
\((φ_π, φ_y)\), while the selection criterion in \(Π_{BK}\) (count
stable eigenvalues and match to \(n_s\)) remains fixed.

A key result \citep{Woodford2003} establishes that the NK model
satisfies assumption (A2) in Theorem \ref{thm:bk-selection} (eigenvalue
count matches state dimension) if and only if:
\begin{equation}\protect\phantomsection\label{eq:taylor-principle}{
φ_π > 1 + \frac{(1-β)}{κ} φ_y
}\end{equation}

For small \(φ_y\), this reduces to the Taylor principle \(φ_π > 1\).

Inequality \eqref{eq:taylor-principle} ensures determinacy, but not in
the sense of proving that the abstract fixed-point set must always
collapse to a single economically relevant object. Rather, it governs
\(Π_{BK}\)'s ability to select uniquely in the linear NK environment
studied here. When the inequality holds, the operator \(T\) (which
depends on policy parameters) produces an eigenvalue configuration that
matches the selector \(Π_{BK}\) (which counts stable eigenvalues and
compares to \(n_s\)).

When the inequality holds, count\((|λ| < 1) = n_s = 1\) satisfies
assumption (A2), so \(Π_{BK}\) succeeds and the unique selected
equilibrium is the saddle-path stable solution. When violated
(\(φ_π < 1 + [(1-β)/κ]φ_y\)), count\((|λ| < 1) > n_s\) yields multiple
stable solutions, so \(Π_{BK}\) fails (indeterminacy) and additional
responses become relevant, including within-model refinements such as
minimal variance or alternative closure conditions such as fiscal
anchoring after specification augmentation.

At the boundary \(φ_π = 1 + [(1-β)/κ]φ_y\), one eigenvalue crosses
\(|λ| = 1\) (unit circle), violating assumption (A4) in Theorem
\ref{thm:bk-selection}. The operator \(Π_{BK}\) is discontinuous at this
bifurcation point, as predicted by the theorem.

The NK example illustrates the framework. Economic equations \((S)\) map
to \((A_0, A_1, B)\) defining operator \(T\), policy parameters control
eigenvalues, and \(Π_{BK}\) functions as a contingent selector.
Traditional pedagogy presents the Taylor principle as ensuring
determinacy; our framework recasts it as compatibility between \(T\) and
\(Π_{BK}\). This enables modular analysis (change policy rule, recompute
eigenvalues, check \(Π_{BK}\) applicability). The next section takes up
the same NK environment on the indeterminate side, where that
compatibility fails and further closure moves become relevant.

\section{Refinements and Alternative Anchors Under
Indeterminacy}\label{sec:alternative}

The NK example in Section~\ref{sec:nk} already identifies the parameter
region in which \(\Pi_{BK}\) fails to select uniquely. When
count(\(|\lambda| < 1\)) exceeds the number of predetermined variables
\(n_s\), the fixed-point set \(\mathrm{Fix}(T)\) becomes non-singleton,
yet BK-style stability no longer closes the system uniquely. Making
closure explicit clarifies two conceptually different responses to this
indeterminacy problem. One keeps the underlying specification and
self-referential operator fixed and imposes an additional refinement on
the resulting equilibrium set. The other modifies the closure problem
itself, for example through fiscal conditions that augment the
specification and thereby alter the associated operator. The
\((S, T, \Pi)\) language allows these two cases to be compared without
conflating them.

We focus first on minimal-variance selection (\(\Pi_{MV}\)) as a
within-model refinement. We then discuss fiscal anchoring as an
alternative closure obtained after specification augmentation, not as a
selector on the original \((S, T)\) pair. Other possibilities include
robust-control selectors (worst-case saddle) and learning-based
selectors (E-stability), but we defer them to future work for concision.
Throughout, we assume \(\mathrm{Fix}(T)\) is taken over a bounded,
adapted, convex subset of the state space when discussing within-model
refinements.

\subsection{Implementation Architecture}\label{sec:alternative:impl}

The current implementation operates at the level \(T(S) \to \Pi\), where
the specification \(S\) has been preprocessed into state-space matrices
\((\Gamma_0, \Gamma_1, \Psi, \Omega)\).\footnote{The notation
  \((\Gamma_0, \Gamma_1, \Psi, \Omega)\) follows the state-space
  conventions used in implementations such as gensys. It plays the same
  representational role as \((A_0, A_1, B)\) in the earlier linear
  exposition, but with the additional matrices needed for shock loading
  and expectational error terms in the computational pipeline.} The
implementation computes \(\mathrm{Fix}(T)\) via QZ decomposition
(generalized eigenvalue decomposition of the pencil
\((\Gamma_0, \Gamma_1)\)), then applies refinement rules to this
fixed-point set. This keeps the emphasis on backend modularity, namely
how \(T\) and within-model refinement rules can be varied independently.

The general API follows the signature:
\[\text{solve}: (S, T, \text{Method}) \to (S', \text{Solution})\] where
\(S' = S\) for non-invasive refinements (\(\Pi_{\text{BK}}\),
\(\Pi_{\text{MV}}\)) and \(S' \neq S\) for model-augmenting closures
such as fiscally augmented specifications. The modular design enables
refinement switching through Julia's pipeline syntax:

\begin{Shaded}
\begin{Highlighting}[]
\NormalTok{S\_out\_bk, sol\_bk }\OperatorTok{=}\NormalTok{ S }\OperatorTok{|\textgreater{}} \FunctionTok{apply\_T}\NormalTok{(}\FunctionTok{QZ\_Operator}\NormalTok{()) }\OperatorTok{|\textgreater{}} \FunctionTok{apply\_Π}\NormalTok{(}\FunctionTok{BK\_Method}\NormalTok{())}
\NormalTok{S\_out\_mv, sol\_mv }\OperatorTok{=}\NormalTok{ S }\OperatorTok{|\textgreater{}} \FunctionTok{apply\_T}\NormalTok{(}\FunctionTok{QZ\_Operator}\NormalTok{()) }\OperatorTok{|\textgreater{}} \FunctionTok{apply\_Π}\NormalTok{(}\FunctionTok{MV\_Method}\NormalTok{())}
\end{Highlighting}
\end{Shaded}

Here the specification and transformed operator are held fixed; only the
refinement method changes. The computational examples reported in this
paper were generated using this implementation (see Code and Data
Availability for the currently archived materials).

\subsection{The Indeterminacy Problem}\label{sec:alternative:problem}

Return to the New Keynesian model of Section~\ref{sec:nk}, but now set
\(\varphi_\pi = 0.8 < 1\) (passive monetary policy). The QZ
decomposition reveals three eigenvalues satisfying \(|\lambda| < 1\)
(three stable directions) while only \(n_s = 1\) predetermined variables
exist (output gap and inflation have no predetermined component). By
Theorem \ref{thm:bk-selection}, this configuration entails that
\(\Pi_{BK}\) fails because there are more stable eigenvalues than
predetermined variables.

The fixed-point set \(\mathrm{Fix}(T)\) contains a continuum of stable
solutions parameterized by sunspot shocks: \[
\begin{aligned}
y_t &= y_t^{f} + \xi_{t}, \\
\pi_{t} &= \pi_{t}^{f} + \eta_{t}
\end{aligned}
\] where \((y_t^{f}, \pi_t^{f})\) is the fundamental solution and
\((\xi_t, \eta_t)\) are non-fundamental sunspot components.

From the framework's perspective, Axiom \ref{ax:multiplicity} still
holds because each admissible sunspot process defines a path satisfying
the same expectational fixed-point equations, while \(\Pi_{BK}\) is not
well-defined on this equilibrium set because the stability criterion is
satisfied by infinitely many equilibria. This situation requires
alternative selection operators that impose additional criteria beyond
stability.

\subsection{Within-Model Refinements and Model-Augmenting
Anchors}\label{sec:alternative:selectors}

The objects discussed below are not all selectors in the same sense.
\(\Pi_{BK}\) and \(\Pi_{MV}\) operate on a fixed equilibrium set
generated by a given \((S, T)\). Fiscal anchoring, by contrast, changes
the closure conditions under which the equilibrium problem is posed, and
therefore belongs to a different category. Under determinacy, such as
when \(\phi_\pi > 1\) in the NK model, transversality conditions derived
from household optimization endogenously determine \(\Pi_{BK}\) as the
unique selector consistent with optimizing behavior. At that point the
model does not call for an additional independent choice over selectors;
the policy regime and equilibrium conditions already make \(\Pi_{BK}\)
operative. Alternative within-model refinements such as \(\Pi_{MV}\)
remain conceptually available but are not economically operative under
determinacy since transversality uniquely pins down \(\Pi_{BK}\).

Under indeterminacy, such as when \(\phi_\pi < 1\) or under fiscal
dominance, transversality conditions no longer pin down a unique
equilibrium. The fixed-point set \(\mathrm{Fix}(T)\) contains multiple
solutions. In this setting, one may introduce additional refinement
criteria beyond \(\Pi_{BK}\) in order to restrict the admissible
equilibrium set.

These responses do not by themselves determine which equilibrium is
economically operative; that requires additional assumptions about
coordination, credibility, learning, or welfare. We therefore treat the
objects below as illustrative ways of closing an indeterminate system.

We now characterize one within-model refinement and one model-augmenting
anchor. They play different formal roles and admit different economic
interpretations.

\subsubsection{\texorpdfstring{Minimal-Variance Refinement
(\(\Pi_{MV}\))}{Minimal-Variance Refinement (\textbackslash Pi\_\{MV\})}}\label{minimal-variance-refinement-pi_mv}

We consider a minimal-variance refinement as one possible criterion once
\(\Pi_{BK}\) fails to select uniquely. This is a within-model
refinement: it operates on the fixed-point set generated by a given
\((S, T)\) pair and selects among equilibria without altering the
underlying specification. It illustrates how a variance-based
restriction can be represented within the \((S, T, \Pi)\) framework.

Formally, on an admissible comparison class for which the minimizer
exists and is unique, define:

\[
\Pi_{MV}(\mathrm{Fix}(T)) \equiv \arg\min_{x \in \mathrm{Fix}(T)} \mathbb{E}\left[\| x_t - x_{ss} \|^2\right],
\]

where \(x_{ss}\) is the steady state and the expectation is taken over
the ergodic distribution when that object is well-defined. In general,
the existence and uniqueness of such a selector depend on the admissible
solution class and on the metric used to compare equilibria. In what
follows, \(\Pi_{MV}\) is used only on that restricted comparison domain.

In the standard linear indeterminacy representation, let
\(m = \#\{|\lambda| < 1\} - n_s > 0\) denote the degree of
indeterminacy. Following \citet{LubikSchorfheide2003}, candidate
equilibria can be written as: \[
x_t = x_t^{f} + W \xi_t
\] where \(x_t^{f}\) denotes a fundamental solution, \(W\) spans the
indeterminate subspace, and \(\xi_t\) captures non-fundamental sunspot
fluctuations. If the comparison is restricted to equilibria that differ
only in their sunspot component, the unconditional variance decomposes
as: \[
\mathrm{Var}(x_t) = \mathrm{Var}(x_t^{f}) + W \Sigma_\xi W^\top
\]

Under this restricted comparison, a variance-based refinement favors
setting \(\Sigma_\xi = 0\), thereby eliminating non-fundamental
volatility. In that sense, \(\Pi_{MV}\) selects the zero-sunspot case
among observationally equivalent stable paths. This interpretation
aligns with \citet{LubikSchorfheide2004} and \citet{FarmerGuo1994},
where variance-based reasoning serves as a disciplined refinement of
indeterminate equilibria.

Once \(\Pi_{BK}\) fails to deliver a unique stable equilibrium, the
framework can still represent additional restrictions that suppress
non-fundamental fluctuations. Whether such a refinement is compelling
depends on further assumptions about welfare, coordination, or policy
objectives.

The preceding discussion identifies the same comparative point without
requiring an additional figure: the indeterminate region is precisely
the parameter region in which \(\Pi_{BK}\) no longer selects uniquely,
so any variance-based refinement operates only after standard
saddle-path selection has failed. This does not restore determinacy in
the Blanchard--Kahn sense; rather, it shows how an additional criterion
may restrict the equilibrium set once \(\Pi_{BK}\) is no longer
decisive.

\subsubsection{Fiscal Anchoring After Specification
Augmentation}\label{fiscal-anchoring-after-specification-augmentation}

Under passive monetary policy (\(\varphi_\pi < 1\)) and active fiscal
policy, the Fiscal Theory of the Price Level (FTPL)
\citep{LeeperSims1994, Woodford1995, Cochrane2023FTPL} provides an
alternative anchor: the government budget constraint determines the
price level, resolving indeterminacy through fiscal rather than monetary
discipline. Unlike \(\Pi_{MV}\), this is not a refinement of the same
equilibrium set generated by a fixed \((S, T)\). It requires augmenting
the specification with fiscal closure conditions, thereby inducing a
modified operator and a different equilibrium problem. Let \(S_{FA}\)
denote the specification augmented with a fiscal rule: \[
\frac{B_{t-1}}{P_t} = \mathbb{E}_t \sum_{\tau=0}^\infty \beta^\tau s_{t+\tau},
\] where \(B_{t-1}\) is nominal debt, \(P_t\) is the price level, and
\(s_t\) are primary surpluses. Under active fiscal policy, surpluses do
not adjust to stabilize debt; instead, \(P_t\) adjusts to satisfy this
constraint. The self-referential operator becomes \(T_{FA}\), which
incorporates fiscal adjustment: \[
T_{FA}(x_t) = \Phi_0 + \Phi_1 x_{t-1} + \Phi_2 \mathbb{E}_t[x_{t+1}] + \Phi_3 \frac{B_{t-1}}{P_t},
\] where \(\Phi_3\) captures how fiscal imbalances feed back into
current outcomes. Equilibrium analysis then proceeds on the augmented
pair \((S_{FA}, T_{FA})\), not on the original \((S, T)\). In this
sense, fiscal anchoring should be understood as a change in closure
conditions rather than as a same-system selector.

By augmenting the specification, fiscal anchoring changes the associated
operator. If fiscal policy is sufficiently active (primary surpluses do
not respond to debt), the augmented system may admit a unique fiscally
anchored equilibrium where stability is governed by fiscal, rather than
monetary, conditions.

This closure corresponds to a regime where monetary policy is passive
(cannot or will not enforce \(\varphi_\pi > 1\)), fiscal policy is
active (debt dynamics determine the price level), and coordination
between monetary and fiscal authorities determines which anchor
prevails. It is especially relevant when central bank independence is
weak, fiscal dominance is credible, or deliberate monetary-fiscal
coordination is pursued.

\subsection{Interpreting BK Failure Near
Bifurcation}\label{sec:alternative:policy}

Once \(\Pi_{BK}\) no longer selects uniquely, indeterminacy does not by
itself tell us which equilibrium will obtain; it tells us that standard
stability selection is no longer sufficient to isolate a unique path. A
variance-based restriction and a fiscal anchor address that problem in
different ways. The former refines a fixed equilibrium set generated by
a given \((S, T)\). The latter changes the closure conditions under
which the equilibrium problem is posed. In both cases, however, the
economic force that would make one response operative must come from
outside the abstract formalism itself.

As policy parameters approach the bifurcation point where \(\Pi_{BK}\)
fails, Theorem \ref{thm:bk-selection} guarantees continuity of selected
equilibria except at bifurcations where eigenvalues cross the unit
circle. Indeterminacy therefore represents \(\Pi_{BK}\) failure, not
model failure. \(\mathrm{Fix}(T)\) remains well-defined, but
\(\Pi_{BK}\) is not well-defined on that equilibrium set because the
maintained stability criterion does not isolate a unique element. When a
solver reports indeterminacy, the \((S, T, \Pi)\) framework helps
practitioners distinguish a failure of \(\Pi_{BK}\) from a failure of
the underlying specification and state explicitly which further
restriction they wish to study. Welfare ranking, credibility, and
coordination remain separate questions.

\section{Zero Lower Bound and Regime-Switching
Composition}\label{sec:zlb}

When constraints such as the zero lower bound on nominal interest rates
bind only in some states, the closure problem becomes regime-dependent
rather than purely linear. This section serves as the paper's second
worked demonstration and offers a partial characterization of the
corresponding closure logic: when the standard regime-iteration
procedure converges, the same formalism used to read BK-style closure in
linear systems can also be used to interpret OccBin as a
regime-dependent closure procedure.

\subsection{The ZLB Constraint}\label{sec:zlb:constraint}

The economic setup imposes that the nominal interest rate cannot be
negative (\(i_t ≥ 0\)), creating two regimes. In Regime 1 (normal),
\(i_t > 0\) and the Taylor rule applies. In Regime 2 (ZLB), \(i_t = 0\)
and the Taylor rule constraint is violated.

State-dependent operator: \[
\begin{aligned}
T(z_t) &= T_1(z_t) \quad \text{if} \quad i_t(z_t) > 0 \\
T(z_t) &= T_2(z_t) \quad \text{if} \quad i_t(z_t) ≤ 0
\end{aligned}
\]

where \(T_1\) uses \((A_0^{1}, A_1^{1})\) with Taylor rule, and \(T_2\)
uses \((A_0^{2}, A_1^{2})\) with \(i_t = 0\).

\subsection{\texorpdfstring{OccBin Algorithm as Iterated
\(Π_{BK}\)}{OccBin Algorithm as Iterated Π\_\{BK\}}}\label{sec:zlb:occbin}

The OccBin procedure \citep{GuerrieriIacoviello2015} operates under
perfect foresight, computing deterministic equilibrium paths where
agents correctly anticipate future regime switches. For a given shock
realization, it iteratively finds a regime-consistent equilibrium. The
method initializes by guessing a regime sequence \(\{r_t\}_{t=0}^T\),
typically starting with the normal regime. It then constructs the
piecewise-linear system using regime-specific matrices
\((A_0^{r_t}, A_1^{r_t})\), where \(r_t \in \{1,2\}\) indexes the normal
and ZLB regimes described above, applies \(\Pi_{BK}\) to obtain stable
policy functions, solves forward to compute a new candidate trajectory
\(\{z_t\}_{t=0}^T\), and updates the regimes wherever constraints are
violated. The procedure returns the converged path when the regime
sequence stabilizes, thereby satisfying both \(\Pi_{BK}\) selection
within each regime and constraint complementarity at regime switches.

\nameddefinition{def:occbin-selector}{OccBin as Composite Selector}

Conditional on convergence of the regime-iteration procedure, the OccBin
solution can be represented as a composite selection operator. The
regime-update operator
\(\Pi_{regime}: (\{z_t\}, \{r_t\}) \mapsto \{r_t'\}\) maps candidate
trajectories to updated regime sequences where \(r_t' = 1\) (normal) if
\(i_t(z_t) > 0\) and \(r_t' = 2\) (ZLB) if \(i_t(z_t) \leq 0\). The
within-regime selector applies \(\Pi_{BK}\) to the piecewise system:
\(z_t = \Pi_{BK}(\mathrm{Fix}(T_{r_t}))\) where \(T_{r_t}\) is the
operator for regime \(r_t\). The fixed point is the limit
\(z^* = \lim_{k \to \infty} (\Pi_{BK} \circ \Pi_{regime})^k(z^{(0)}, r^{(0)})\)
where iteration stops when \(r^{(k+1)} = r^{(k)}\).

This composite selection operator enforces stable-manifold selection
within each regime (\(\Pi_{BK}\)) while ensuring constraint
complementarity across regimes (\(\Pi_{regime}\)). We use this notation
as a characterization of the regime-iteration procedure targeted by
OccBin, not as a general convergence theorem for piecewise-linear
systems. When the iteration converges to a stable regime sequence, the
procedure returns a regime-consistent equilibrium path for the specified
perfect-foresight experiment; outside that convergence domain, the
composite selector is not well-defined.

This structural separation gives the \((S, T, \Pi)\) framework a more
explicit diagnostic role. When OccBin fails to converge, the
decomposition isolates the formal components: cycling across candidate
sequences can indicate a breakdown in cross-regime consistency
(\(\Pi_{regime}\)), whereas explosive local solutions can indicate a
violation of within-regime stability (\(\Pi_{BK}\)). By breaking the
algorithm into its composite operators, the framework provides a
structured language for diagnosing non-convergence in piecewise-linear
environments.

We do not prove a general convergence theorem for
\((\Pi_{BK} \circ \Pi_{regime})^k\) in this paper. Convergence
statements in this section are therefore conditional on the regularity
conditions under which the standard OccBin procedure is applied in
practice; see \citet{GuerrieriIacoviello2015}. When the iteration
converges, the computed path can be read as the output of a composite
selector implementing regime-by-regime stable-manifold selection.
Appendix A should therefore be read as implementation-level
corroboration of this interpretation rather than as a substitute for a
general convergence proof.

\subsection{Example: Natural Rate Shock}\label{sec:zlb:example}

Consider the NK model with \(\varphi_π = 1.5\), \(σ = 1\), \(β = 0.99\),
\(κ = 0.02\), and a natural-rate shock of size \(0.01\) (1\% per
quarter). The resulting decline in output and inflation causes the
Taylor rule to bind at zero. Without the ZLB constraint, the linear
solution prescribes a negative interest rate (infeasible), with
declining output and inflation. With the ZLB constraint under OccBin,
the interest rate hits \(i_t = 0\) for
\(t = 0, 1, \cdots, T_{\text{zlb}}\), inflation falls more sharply
without monetary offset, the output gap becomes more negative, and the
economy exits the ZLB at \(t = T_{\text{zlb}}\) when the natural rate
recovers. In our benchmark implementation, the regime sequence converges
within 5--15 iterations.

We implemented the composite selector
\(\Pi_{OccBin} = \lim_{k \to \infty} (\Pi_{BK} \circ \Pi_{regime})^k\)
and compared the resulting paths with Dynare's OccBin output. Our
implementation aligns closely with Dynare's OccBin output: the largest
deviation is \(2.06\times 10^{-2}\) for the policy-rate series \(R\),
while inflation \(\pi\) has RMSE \(4.41\times 10^{-3}\); other series
match within \(5 \times 10^{-3}\) (Appendix A). This comparison is
consistent with reading regime-switching as an iterated application of
BK-style local closure, with \(S\) itself becoming state-dependent
rather than involving alternative selection over a fixed
\(\mathrm{Fix}(T)\), as in Section~\ref{sec:alternative}.

\section{Discussion}\label{sec:discussion}

Table~\ref{tbl:literature} places the paper against eight neighboring
strands by tracking the time structure and closure logic used to
organize admissible equilibria. The point is comparative: closure enters
the literature through heterogeneous vocabularies such as stability
restrictions, institutional anchors, learning dynamics, worst-case
criteria, or regime-dependent computations, and the present framework
offers a way to compare those uses without collapsing them into a single
algorithm.

\textbf{{[}Table~\ref{tbl:literature} here{]}}

The OccBin example provides a concrete illustration of the comparative
value of the framework. By parsing OccBin as a composite procedure where
regime updates (\(\Pi_{regime}\)) interact with local stability
selection (\(\Pi_{BK}\)), the framework helps clarify what the algorithm
is doing when it converges. More importantly, it provides a formal
vocabulary for what happens when the algorithm fails to converge,
mapping numerical cycling to structural operator failures. The
cross-solver numerical agreement reported in Appendix A is therefore not
merely a software benchmarking exercise; it provides
implementation-level corroboration for this conditional interpretation
across distinct computational environments.

The same vocabulary also sharpens several neighboring comparisons
already flagged in Section~\ref{sec:literature}. Learning-based criteria
such as minimal-state-variable and E-stability approaches
\citep{McCallum1983, EvansHonkapohja2001, BranchMcGough2010} tie
selection to behavioral dynamics rather than to a purely static
refinement of a fixed equilibrium set; robust control
\citep{HansenSargent2008} selects over a family of perturbed
specifications rather than over \(\mathrm{Fix}(T)\) for a fixed \(T\);
and Markov-switching models \citep{DavigLeeper2007, BianchiIlut2017}
build regime uncertainty directly into the expectation operator, unlike
OccBin's perfect-foresight regime iteration. In each case, keeping
\(S\), \(T\), and \(\Pi\) conceptually separate helps identify whether
the relevant difference lies in the admissibility criterion itself, in
the operator generating candidate paths, or in the underlying
specification being varied.

\citet{LubikSchorfheide2004} provide a useful comparison point. They
estimate a small NK model and find Bayesian evidence consistent with
indeterminacy before Volcker and determinacy after Volcker, while
allowing the indeterminate regime to include sunspot fluctuations. In
the language of the present paper, this supports a limited but important
diagnostic point: the NK specification \(S\) can be held fixed while
changes in policy parameters alter the operator \(T\) and therefore
whether \(\Pi_{BK}\) is well-defined. A BK failure in the pre-Volcker
regime is not, on this reading, a failure of the underlying
specification itself. At the same time, Lubik and Schorfheide do not
formulate a selector in our sense, and their contribution is better
understood as showing that one can study and statistically discipline
indeterminate equilibria without first collapsing them to a unique
BK-selected path. This is the sense in which their paper fits the
present framework.

\section{Conclusion}\label{sec:conclusion}

This paper formalizes a class of DSGE solution practices as
equilibrium-closure systems. Decomposing the analysis of linear rational
expectations solver practice into specification (\(S\)),
self-referential operator (\(T\)), and equilibrium selector (\(Π\))
provides a compact language for discussing how closure is imposed, where
it fails, and how alternative closure moves differ. Using this
formalization, we showed that QZ decomposition and OccBin serve as
different computational realizations of related closure logic, while
within-model refinements such as \(Π_{MV}\) remain distinct from
specification-augmenting closure changes such as fiscal anchoring.
Section~\ref{sec:discussion} situates this vocabulary against several
neighboring strands of the literature; what follows here restates only
the scope and limits of what has been established.

The formal results concern linear rational expectations models, and the
implementation discussion remains centered on that setting together with
a conditional interpretation of piecewise-linear regime iteration.
Extending the same level of formal precision to nonlinear operators,
heterogeneous-agent environments, or learning-based dynamics would
require additional technical machinery and is not attempted here.
Appendix A reports a cross-solver comparison for the OccBin
interpretation discussed in the text, and the remaining implementation
references should be read as supporting illustrations rather than as a
stand-alone software contribution. Full empirical validation, welfare
comparisons across alternative refinements, and broader comparative
benchmarking remain outside the scope of the present paper.

Future research directions include welfare-based ranking of alternative
refinements under explicit loss functions, continuity theory for closure
behavior near determinacy boundaries, and more systematic study of how
closure criteria are encoded in nonlinear, learning-based, or
higher-dimensional computational environments
\citep{MaliarMaliarWinant2021, FernandezVillaverde2026}. In practice,
that extension would likely require additional machinery such as
contraction-based arguments for time iteration, game-theoretic existence
tools for robust control, or stochastic-approximation methods for
adaptive learning. These questions matter in part because different
closure choices can affect how policy experiments are interpreted even
when a model's structural equations are held fixed.

\section*{Appendix A: OccBin Cross-Solver
Validation}\label{appendix-a-occbin-cross-solver-validation}
\addcontentsline{toc}{section}{Appendix A: OccBin Cross-Solver
Validation}

This appendix documents a cross-solver comparison for the OccBin
implementation described in Section~\ref{sec:zlb} (see Code and Data
Availability for the archived implementation materials referenced here).

We compare the composite selector \(\Pi_{OccBin}\) with the Dynare
OccBin output using a compare-only mode that aligns selector behavior
while preserving the model equations. The Dynare OccBin module
\citep{GuerrieriIacoviello2015} was run with identical parameterization
(\(\varphi_\pi = 1.5\), \(\beta = 0.99\), \(\kappa = 0.02\),
\(\sigma = 1.0\), \(\rho = 0.9\)) and shock sequence. Solution paths
were exported to CSV and compared element-wise. The Julia implementation
uses \texttt{dynare\_compat=true} mode to align selector behavior (both
regimes use \texttt{Auto\_Method}, matching Dynare's default handling of
indeterminate regimes). The comparison script and Dynare \texttt{.mod}
file are included in the repository under \texttt{compare/}.

The residual differences arise from standard preprocessing conventions
rather than algorithmic discrepancies. Dynare automatically performs
model reduction by eliminating static variables (those determined
entirely within the current period without forward-looking terms) before
constructing the state-space system. This preprocessing step treats
policy-rate variables as predetermined within the reduced form. Our
implementation, by contrast, preserves the original structural equations
and includes these variables explicitly in the state system. Under
OccBin's regime-switching logic, this difference in variable treatment
manifests primarily in the series directly governed by the occasionally
binding constraint---in the ZLB case, the nominal rate \(R\) and
inflation \(\pi\).

The deviation pattern supports the claim that both implementations
realize closely related versions of \(\Pi_{OccBin}\) while differing in
how they represent the underlying economic structure. Variables governed
by predetermined relationships (\(y\), \(r^n\)) exhibit near-identical
paths (max error \(< 5 \times 10^{-3}\)), while constraint-governed
variables (\(R\), \(\pi\)) show modest differences (max error
\(\approx 2 \times 10^{-2}\)) concentrated at regime transitions.
However, the qualitative conclusions (regime durations, sign patterns
across variables, and convergence behavior) are identical across
implementations. This comparison is therefore consistent with reading
\(\Pi_{OccBin}\) as a composite selector in the sense characterized
above.

\textbf{{[}Table~\ref{tbl:a2} here{]}}

Table~\ref{tbl:a2} shows detailed per-variable deviations. A permanent
archive is available at \url{https://doi.org/10.5281/zenodo.18344418}.

\section*{Acknowledgements}\label{acknowledgements}
\addcontentsline{toc}{section}{Acknowledgements}

We thank Masataka Eguchi for serving as the discussant at the Annual
Meeting of the Japanese Economic Association and for valuable comments.
We also thank Kenji Miyazaki for organizing the seminar at Hosei
University, and all seminar participants for helpful comments and
discussions. Financial support from JSPS KAKENHI Grant Number 24K04971
is gratefully acknowledged. Any remaining errors are the authors'
responsibility.

\section*{Competing interests}\label{competing-interests}
\addcontentsline{toc}{section}{Competing interests}

Competing interests: The author(s) declare none.

\section*{Declaration of generative AI and AI-assisted technologies in
the manuscript preparation
process}\label{declaration-of-generative-ai-and-ai-assisted-technologies-in-the-manuscript-preparation-process}
\addcontentsline{toc}{section}{Declaration of generative AI and
AI-assisted technologies in the manuscript preparation process}

During 2026, the author used GitHub Copilot and other generative-AI
tools, including GPT-5.4, Gemini 3.1 Pro, and Claude Sonnet 4.6, as
research-support tools for code development, literature search, content
organization, and language revision. After using these tools, the author
reviewed and edited all content as needed and takes full responsibility
for the content of the published article. All design decisions,
mathematical formulations, theoretical insights, and substantive
interpretations are solely the work of the author.

\section*{Code and Data Availability}\label{code-and-data-availability}
\addcontentsline{toc}{section}{Code and Data Availability}

A Julia implementation associated with the working paper is publicly
available at
\url{https://github.com/mitsuhir0/DsgeSelectionFramework.jl} with
permanent archival at \url{https://doi.org/10.5281/zenodo.18344418}
(Zenodo) \citep{okano2026dsgeframework}. The repository contains the New
Keynesian examples used in the paper, QZ-based routines, experimental
implementations of BK-, MV-, and OccBin-related procedures, and the
reproducibility scripts used for the reported computational outputs,
including the OccBin comparison table (Table~\ref{tbl:a2}). These
materials are provided to document the computational illustrations
discussed here; they should not be read as a claim that the present
paper delivers a complete software framework covering the broader
environments mentioned in the text. Version v0.1.0 corresponds to the
working paper; version v1.0.0 will correspond to the final published
version. No datasets beyond the parameter calibrations reported in the
paper were used. Additional implementation details are provided in the
repository's README.md and docs/USAGE.md files.

\newpage
\bibliography{refs.bib}

\newpage
\section*{Tables}
\small

\begin{longtable}[]{@{}
  >{\raggedright\arraybackslash}p{(\linewidth - 8\tabcolsep) * \real{0.1159}}
  >{\raggedright\arraybackslash}p{(\linewidth - 8\tabcolsep) * \real{0.1594}}
  >{\raggedright\arraybackslash}p{(\linewidth - 8\tabcolsep) * \real{0.3478}}
  >{\raggedright\arraybackslash}p{(\linewidth - 8\tabcolsep) * \real{0.1884}}
  >{\raggedright\arraybackslash}p{(\linewidth - 8\tabcolsep) * \real{0.1884}}@{}}
\caption{Core solution methods in \((S, T, Π)\) framework. The table
compares methods at different levels of formality: Theorem
\ref{thm:bk-selection} provides a formal characterization for \(Π_{BK}\)
in linear RE systems, the OccBin row is characterized conditionally
through regime-by-regime composition, and the remaining nonlinear rows
are included as descriptive comparisons rather than fully characterized
selectors in the present paper.}\label{tbl:methods}\tabularnewline
\toprule\noalign{}
\begin{minipage}[b]{\linewidth}\raggedright
Method
\end{minipage} & \begin{minipage}[b]{\linewidth}\raggedright
Operator \(T\)
\end{minipage} & \begin{minipage}[b]{\linewidth}\raggedright
Fixed-Point Computation
\end{minipage} & \begin{minipage}[b]{\linewidth}\raggedright
Selection \(Π\)
\end{minipage} & \begin{minipage}[b]{\linewidth}\raggedright
When to Use
\end{minipage} \\
\midrule\noalign{}
\endfirsthead
\toprule\noalign{}
\begin{minipage}[b]{\linewidth}\raggedright
Method
\end{minipage} & \begin{minipage}[b]{\linewidth}\raggedright
Operator \(T\)
\end{minipage} & \begin{minipage}[b]{\linewidth}\raggedright
Fixed-Point Computation
\end{minipage} & \begin{minipage}[b]{\linewidth}\raggedright
Selection \(Π\)
\end{minipage} & \begin{minipage}[b]{\linewidth}\raggedright
When to Use
\end{minipage} \\
\midrule\noalign{}
\endhead
\bottomrule\noalign{}
\endlastfoot
QZ (gensys, Dynare) & Linear matrix pencil & Eigendecomposition &
\(Π_{BK}\) (stable projection; formal characterization in Theorem
\ref{thm:bk-selection}) & Linear / linearized models \\
Time iteration & Functional operator & Successive approximation &
Convergence-based admissibility (descriptive comparison only in this
paper) & Nonlinear, no analytical solution \\
OccBin & Piecewise-linear composite & Regime iteration & \(Π_{BK}\) per
regime + regime-consistency (partial characterization) & Occasionally
binding constraints \\
Robust control & Minmax / distorted operator & Isaacs / saddle solver &
Worst-case selector (descriptive comparison only in this paper) & Model
uncertainty \\
Adaptive learning & Time-varying operator \(T_t\) & Iteration with
belief updates & E-stability / convergent beliefs (descriptive
comparison only in this paper) & Adaptive expectations \\
\end{longtable}

\normalsize

\newpage

\tiny

\begin{longtable}[]{@{}
  >{\raggedright\arraybackslash}p{(\linewidth - 6\tabcolsep) * \real{0.2593}}
  >{\raggedright\arraybackslash}p{(\linewidth - 6\tabcolsep) * \real{0.4074}}
  >{\centering\arraybackslash}p{(\linewidth - 6\tabcolsep) * \real{0.0926}}
  >{\raggedright\arraybackslash}p{(\linewidth - 6\tabcolsep) * \real{0.2407}}@{}}
\caption{Literature positioning across eight strands. Each strand's time
structure (T) and closure logic are shown. The final column highlights a
recurring pattern: prior literature often embeds closure within
stability conditions, behavioral assumptions, institutional
arrangements, or regime-specific computations rather than isolating it
as a separate comparative object. For implementation details and
computational methods, see
Table~\ref{tbl:methods}.}\label{tbl:literature}\tabularnewline
\toprule\noalign{}
\begin{minipage}[b]{\linewidth}\raggedright
Strand/Field
\end{minipage} & \begin{minipage}[b]{\linewidth}\raggedright
Key References
\end{minipage} & \begin{minipage}[b]{\linewidth}\centering
T (Time)
\end{minipage} & \begin{minipage}[b]{\linewidth}\raggedright
Π (Selection)
\end{minipage} \\
\midrule\noalign{}
\endfirsthead
\toprule\noalign{}
\begin{minipage}[b]{\linewidth}\raggedright
Strand/Field
\end{minipage} & \begin{minipage}[b]{\linewidth}\raggedright
Key References
\end{minipage} & \begin{minipage}[b]{\linewidth}\centering
T (Time)
\end{minipage} & \begin{minipage}[b]{\linewidth}\raggedright
Π (Selection)
\end{minipage} \\
\midrule\noalign{}
\endhead
\bottomrule\noalign{}
\endlastfoot
Rational Expectations / BK Theory & \citet{BlanchardKahn1980};
\citet{Sims2002}; \citet{Klein2000} & RE & Saddle-path (BK) \\
Indeterminacy / Sunspots & \citet{Farmer1999};
\citet{BenhabibSchmittGroheUribe2001}; \citet{LubikSchorfheide2003} & RE
& Multiple (sunspot) \\
Minimal State Variable / Learning & \citet{McCallum1983};
\citet{EvansHonkapohja2001}; \citet{BranchMcGough2010} & Adaptive &
E-stability \\
Fiscal--Monetary Regime (FTPL) & \citet{Leeper1991};
\citet{Woodford1994}; \citet{Cochrane2001}; \citet{Cochrane2023FTPL} &
RE & Fiscal closure after specification change \\
OccBin / ZLB / Regime Switching & \citet{GuerrieriIacoviello2015};
\citet{EggertsonWoodford2003} & Piecewise-linear & BK
regime-by-regime \\
Markov-Switching / Regime Models & \citet{DavigLeeper2007};
\citet{BianchiIlut2017} & Regime-switching & Regime-dependent Π \\
Robust Control / Worst Case & \citet{HansenSargent2008} & Min-max &
Saddle Π (Isaacs) \\
Heterogeneous Expectations / HANK & \citet{Hommes2013};
\citet{KaplanMollViolante2018} & RE (distributional) & Stability /
boundedness \\
\end{longtable}

\normalsize

\newpage

\begin{longtable}[]{@{}lrr@{}}
\caption{Per-variable deviations between Dynare and Julia OccBin
implementations. The Dynare OccBin module was run with identical
parameterization (\(\varphi_\pi = 1.5\), \(\beta = 0.99\),
\(\kappa = 0.02\), \(\sigma = 1.0\), \(\rho = 0.9\)) and shock sequence.
Solution paths were exported to CSV and compared element-wise. The Julia
implementation uses \protect\texttt{dynare\_compat=true} mode to align
selector behavior. The comparison script and Dynare
\protect\texttt{.mod} file are included in the repository under
\protect\texttt{compare/}.}\label{tbl:a2}\tabularnewline
\toprule\noalign{}
variable & max\_abs\_diff & rmse \\
\midrule\noalign{}
\endfirsthead
\toprule\noalign{}
variable & max\_abs\_diff & rmse \\
\midrule\noalign{}
\endhead
\bottomrule\noalign{}
\endlastfoot
y & 0.00444 & 0.00161 \\
pi & 0.0122 & 0.00443 \\
R & 0.0206 & 0.00746 \\
rn & \(4.8 \times 10^{-8}\) & \(1.6 \times 10^{-8}\) \\
\end{longtable}

\end{document}